# BEHAVIOR OF SELF-MOTIVATED AGENTS IN COMPLEX NETWORKS

**Sundong Kim**
Dept. of Industrial and Systems Engineering
KAIST, Daejeon, South Korea

**Jin-Jae Lee**
Dept. of Physics
KAIST, Daejeon, South Korea

## ABSTRACT

Traditional evolutionary game theory describes how certain strategy spreads throughout the system where individual player imitates the most successful strategy among its neighborhood. Accordingly, player doesn't have own authority to change their state. However in the human society, peoples do not just follow strategies of other people, they choose their own strategy. In order to see the decision of each agent in timely basis and differentiate between network structures, we conducted multi-agent based modeling and simulation. In this paper, agent can decide its own strategy by payoff comparison and we name this agent as "Self-motivated agent". To explain the behavior of self-motivated agent, prisoner's dilemma game with cooperator, defector, loner and punisher are considered as an illustrative example. We performed simulation by differentiating participation rate, mutation rate and the degree of network, and found the special coexisting conditions.

## 1    INTRODUCTION

### 1.1    Overview

Traditional evolutionary game describes how successful strategy spreads in population(Jörgen M. Weibull., 1995). Depending on the payoff structure, agent follows the strategy of most successful neighbor. In that way the whole system adopt a certain strategy. But human society is more complicated that it is not explained by only adopting strategy. Still, there are many peoples trying to follow how other people live, but not everyone makes a decision by mimicking their role model. For example, by only using adopting strategy, we can't explain how great politician grew up from bed neighborhood. Of course there is more probability that he/she will be up to he/she's neck in crime. But he/she might have dreamed of changing the world, or had more advantage to get a good score while he/she is in bed neighborhood. Previous researches hadn't focused on this characteristic of agent so that it is not a sufficient model to explain any autonomous society.

In this paper, we suggest the decision process of "Self-motivated agent". Self-motivated agent can guess which types of strategies are available regardless of their neighborhood, and it can voluntarily choose its strategy by comparing payoff and change its state by itself. Participants play prisoner's dilemma game with their neighborhood and find the maximum payoff that they can get, and finally change to relevant state.

In our model, there are four types of player - cooperator, defector, loner and punisher. To explain society more accurately, we added loner and punisher into traditional prisoner's dilemma game. Loner, an agent with indifferent mind, gets fixed payoff(Dawes, 1980), and punisher is an agent introduced to penalize defector. (Fehr & Gächter, 2002; Rockenbach & Milinski, 2006). And we also considered other three factors, participation rate and mutation rate, and network structure.

Since every agent makes their decision at different time, we can say that only some of the players join the game on each time. To model this factor, we suggest participation rate to set how many players will participate in a time period. Surprisingly, this probability not only controls the speed of dynamics, but also becomes the important factor of deciding the overall behavior of agents. And mutation is also adopted in our model (Glenn W. Rowe, 1985). It corresponds to mutation in the DNA, or in cultural evolution it describes individuals experimenting with new behavior. Sometimes human-being also makes decision by throwing dice, which means they leave it to chance. Mutation is happened regardless of whether the agent is a participant or not.

Complex network structure is also considered (S.Boccaletti et al., 2005). Six different networks are used in our paper, and we found different coexisting condition by changing network degree and structure.

## 1.2   Previous Research

As a famous example of game theory, prisoner's dilemma is a kind of non-zero-sum game where two people participate. If both players cooperate, they will get the largest benefit together. However, if both players choose selfish strategy to maximize personal gain, then they can't get any profit and hence it turns out to be the dilemma. Even there are a lot of people, not two people with two-strategy of cooperators and defectors, we know that defectors prevail through replicator dynamics (Hofbauer & Sigmund, 1998). This result depicts our sociology (Dawes, 1980).

From the point of view of evolutionary theory however, there is a problem that defectors prevail. Robert Axelord and William. D. Hamilton for the first time to solve these problems by using this game theory (Axelrod & Hamilton, 1981). This is just a matter of evolution theory and involve a lot of social issues(Fehr & Fischbacher, 2003; Wedekind & Milinski, 2000). In this way, the effort what you looking for that cooperator can emerge and persists through game theory is said Evolutionary Game Theory.

In 1992, Martin A. Nowak and Robert M. May introduced a spatial structure (Nowak & May, 1992). Individuals are located in the normal grid, and collect the aggregate rewards of interaction with neighbors to mimic the strategy of the most successful neighbor. Studies on attributes about spatial structure in general two-strategy game was carried (Ch Hauert, 2002). Additionally, in snow drift game, the effect of spatial structure gave birth to other result (Christoph Hauert & Doebeli, 2004). Instead of these two-strategy games, they have also been studying games that some elements have been added.

In biology, the phenomenon that three-strategy coexist have been founded (Zamudio & Sinervo, 2000). And these systems have been simulated in rock-paper-scissors game (Kerr, Riley, Feldman, & Bohannan, 2002). From the point of view of evolutionary game theory, individual can follow a strategy of cooperators, defectors, and loners (Dawes, 1980). Regardless of neighbors, loners have the fixed payoff, hence they get more payoff than defectors. On the other hand, they don't get benefit by participating in the game, so cooperators prevail around loners. Therefore, strategy of these three types is cyclic dominant, so they coexist and overcome the dilemma (Hauert et al., 2002). Additionally, the concept of punisher is suggested. Punisher has the opportunity to punish and imposes to co defecting co-players but these actions are costly (Fehr & Gächter, 2002; Rockenbach & Milinski, 2006).

In 2009, the concept of mutation in the game of four strategies, cooperators, defectors, loners, and punishers was introduced (Traulsen et al., 2009). A mutation occurs with probability mutation rate and a mutant switches to a different random strategy. The dynamics has changed as the mutation rate increases. Change of dynamic describes the cultural evolution differs from the genetic reproduction which has rare mutations.

Previous researches have been carried out on lattice structures. More recently, interest in which lattice structure is extended by graphs and social networks has increased and it also affects evolutionary game theory. In the beginning, there is a research about evolutionary dynamics on graphs (Lieberman et al., 2005). Each vertex of graph represents an individual and weighted edges of graph represent reproductive rate. The result shows that amplifying random drift or selection in evolutionary dynamics is determined by forms of graphs. In addition, it said how to access evolutionary games on graphs.

Along this study, the critical point to prevail cooperators was founded by analytical and numerical simulations in two-strategy game of cooperators and defectors on several graphs and networks (Ohtsuki, Hauert, Lieberman, & Nowak, 2006). This system followed 'death-birth' process: in each time step, a random individual is chosen to die, and the neighbors compete for the empty site proportional to their fit-ness. Also, previous research demonstrated that agent's adaptive expectation plays an important role in cooperation emergence on complex networks (Ohtsuki et al., 2006).

## 2 MODEL

### 2.1 Strategies of Agent

In our game, four different players exist. Cooperator is an agent that gives benefit b to all players except loner, but to be a cooperator it has to pay cost c for neighboring players except loner. Cooperator also gets benefit by neighboring cooperator. Defector also gets benefit by neighboring cooperator without paying any cost, but it gets certain amount of penalty β from neighboring punisher. In the simple game with these two types of agent, defector always has advantages over cooperator (Hofbauer & Sigmund, 1998). Loner is an agent who is not interested in this game, so it gets fixed payoff σ regardless of having any types of neighbor. Becoming loner is the best strategy if there is no cooperator around, but if there exists any cooperator neighbor agent will not choose to be a loner directly. At least it has to calculate different payoff functions if we assume total payoff of loner is bigger than benefit. Loner's relevant payoff $\sigma$ between its opponent is set to be $0 < \sigma < b$. Punisher is an agent introduced to penalize defector like police officer. In our paper, punisher is not a type of cooperator, their function is just giving punishment to neighboring defectors. It is also benefited by neighboring cooperator but it has to bear small penalty γ ($< \beta$) to penalize neighboring defector. Comparing cost of altruism, penalty of defect and small penalty to penalize, c $\ll \gamma < \beta$ is established. Because each cooperator has to pay maximum cost d × c by considering network degree of d, and it has to be smaller than γ and $\beta$. Otherwise agent will not choose to do altruistic behavior. We brought the concept of punisher and loner from previous research (Traulsen et al., 2009), but we put minor change of their role and payoff functions – in our model punisher doesn't act as a cooperator, and cooperator doesn't pay costs for neighboring loner .

### 2.2 4-Person prisoner's dilemma game

Let's consider people doing teamwork. We can divide people's behavior into four different types. People who trust other people and shows altruism – cooperator, free-rider who only gets benefit from teammate – defector, person who is not interested in or not joining that work – loner, person who criticize bad teammate, but gets bed reputation because he always lays out a sermon – punisher (Greenberg J., & Baron, R.A, 1997). Agent in our model acts exactly as these four different types of people and decide their role by playing 4-person prisoner's dilemma game with their neighbors. And the result will show how different behavior agent shows according to change of parameter and network structure.

The model consists of N agents playing 4-person prisoner's dilemma game. Agents are located on the vertices of network. In every simulation step, agents play game with other agents within their own neighborhood. The agent can choose four strategies: cooperation, defection, indifferent, or punish. For example if player A is defector and player B is punisher, punisher give penalty to defector, and it also get minus payoff while penalizing, so their payoff for this game is (−γ , −β). The payoff matrix is as follows :

|  | Cooperator | Defector | Loner | Punisher |
|---|---|---|---|---|
| Cooperator | ( b – c , b – c ) | (−c , <u>b</u> ) | (−c , σ ) | (−c , <u>b</u> ) |
| Defector | ( <u>b</u> , – c ) | ( 0 , 0 ) | ( 0 , <u>σ</u> ) | (−β , −γ ) |
| Loner | ( σ , – c ) | ( <u>σ</u> , 0 ) | ( **<u>σ</u>** , **<u>σ</u>** ) | ( <u>σ</u> , 0 ) |
| Punisher | ( <u>b</u> , – c ) | ( −γ , −β) | ( 0 , <u>σ</u> ) | ( 0 , 0 ) |

Table 1 : 4-person prisoner's dilemma game used in our model

By adding loner and punisher from the traditional prisoner's dilemma game, dominant strategy is to choose loner for both players. So assuming that the agent plays prisoner's dilemma game one by one with its neighbor, we can estimate that all players eventually become loner. But if we consider the society of degree more than 1, agent has to deal with multiple neighbors. Present paper, we assume that each agent knows the strategy of their neighbors. Since each agent is facing with their neighborhoods in the game, it gathers all information of their neighborhoods states. And calculate how much total payoff it can get if it changes into certain state. Detailed explanation is in the next section.

## 2.3   Update Mechanism

To see the dynamics of transition and find different behavior, we added participation probability on our model. If participation rate is 1, every player in the game changes their state each time. In each step, participants are randomly selected depending on participation rate, and every participating agent calculates their expected payoff using the information of their neighbor. And player finally changes its state that guarantees maximum payoff. If multiple maximum payoff value exists, agent picks one state randomly. In our simulation, small random number ε is added to the payoff function so that agent can pick the maximum payoff between four.

We only considered agent's first-step neighbor to get payoff function. For example, there is an agent calculating their options and decide to be a defector. If there is a punisher on its neighbor and whether punisher has another defector on their neighborhood or not, penalty value for that defector doesn't change.

The number of corresponding neighborhood – cooperator, defector, loner, punisher is counted as $n_c, n_d, n_l, n_p$. For example, the payoff function of cooperator can be evaluated as $n_c\, b - (n_c + n_d + n_p)\, c$ because it get benefits from $n_c$ neighboring cooperators and pay costs c to $(n_c + n_d + n_p)$ neighbors for altruistic acts.

The payoff and final strategy can be described as follows :

$$Payoff_{coorporator} = n_c \times b - (n_c + n_d + n_p) \times c$$

$$Payoff_{defector} = n_c \times b - n_p \times \beta$$

$$Payoff_{loner} = (n_c + n_d + n_l + n_p) \times \sigma$$

$$Payoff_{punisher} = n_c \times b - n_d \times \gamma$$

$$\text{Strategy } S = \arg\max_{S} Payoff_s$$

## 2.4   Spatial Structure

Six types of complex networks are used in this paper. Lattice networks of Moore-neighborhood (Network degree d = 8) are used as a baseline to see the coexisting behavior depending on participation rate and mutation rate. And lattice network of von-Neumann neighborhood(d=4) and network with d=16 are used to compare how degree change affects behavior of agents.

And Cellular(Rives, A. W., & Galitski, T. 2003), Core-Periphery(Borgatti, S. P., & Everett, M. G. 2000), Erdős-Rényi,(Erdős, P., & Rényi, A. 1960), Scale-Free(Barabási , A. –L., & Albert, R. 1999), Small-World(Watts, D. J., & Strogatz S. H. 1998) network are used. These networks are complex networks on the ground of having heterogeneous network degree for each node.

Previous research has proven that there are different critical points for cooperators to prevail over the several types of complex networks. (Ohtsuki et al., 2006). And cooperation frequency fluctuations in a Barabási and Albert (BA) network and Watts and Strogatz (WS) small world networks are compared (Bo.X, 2012). In this paper, we found different coexisting points and interesting phenomena depending on network structure and degree.

## 2.5 Participation rate and Mutation rate

Participation rate p and mutation rate μ plays the important role in our simulation. In our setting, participation rate decides how many agents decide to change their state on each time and mutation rate describes individuals experimenting with new behaviors. In our model, randomly selected participants play game first, and mutation occurs with probability μ in each update step. Previous study suggests that adjusting mutation rate can result significant change in the societies (Traulsen et al., 2009), and we found different stationary states depending on participation rate and mutation rate.

## 3 RESULT & DISCUSSION

## 3.1 Simulation Setting

Initially, 2,500 agents are located on the grid space, and four types of agents are randomly distributed throughout the space. The degree of the network is 8 by querying Moore neighborhood. For visualizing, we used grid space with world height 50 and width 50, and the bias is removed by using torus space. Participation rate p and mutation rate μ is varied in the interval [0, 1] and benefit b, cost c, penalty $\beta$, small penalty γ is set to be 100, 5, 150, 50 respectively, And loner's total payoff for this game is set to be b for Moore neighborhood. ($\sigma = \frac{b}{8}$). Since initial place of agent located can affect the final coexisting probability, we simulated 30 times by changing random seeds. In our simulation, we simulated by controlling participation rate and mutation rate together by batch run. We picked 4 participation rate p = 0.0001, 0.01, 0.1, 1 by changing mutation rate μ = 0, 0.0001, 0.001, 0.01, 0.01. Our null hypothesis is when p = 1, μ = 0. Table 2 and 3 shows the simulation parameters and virtual experiment design.

| Figure | Value |
|---:|---|
| Number Of Agent | 2500 |
| Space Height | 50 |
| Space Width | 50 |
| Benefit (b) | 100 |
| Cost (c) | 5 |
| Penalty (β) | 50 |
| Small Penalty (γ) | 15 |
| Loner's payoff (σ) | 12.5 |

Table 2 : Simulation parameters

| Experiment Variable name | Experiment Design | Implication |
|---|---|---|
| Participation rate (p) | 0.001, 0.01, 0.1, 1(4 cases) | Participation rate of each agent |
| Mutation rate (μ) | 0, 0.0001, 0.001, 0.01, 0.1 (5 cases) | Randomly mutated rate |
| Simulation ending time | 10000 ticks | Stable point for simulation |
| Each experiment is replicated 30 times by changing random seeds. | | |

Table 3 : Virtual experiment design of scenario of interests

## 3.2  Participation rate & Mutation rate

Participation and Mutation rate are important factors that decide coexisting condition. We simulated 4-agent prisoner's dilemma game by controlling Participation rate and Mutation rate. Figure 1 describes simple example of how coexisting probability and behavior changes depending on the participation rate and mutation rate. Red, blue, green, black dots are cooperator, defector, loner, and punisher agents. Left is a stable situation with p = 1, μ = 0. Middle is loner-dominant society after changing mutation rate to μ = 0.01. Right you can see coexistence again by changing p = 0.1. We can say that stable structure is collapsed by adopting mutation, because cooperators who give benefit have a chance to change into other agents and their neighborhoods decide to change themselves in the next step. But by put participation rate into our simulation, we can slow down the second part of the action and give more chance to survive the coexisting structure.
(Fixed parameters : 2,500 agents in grid space with Moore neighborhood, b=100, c=5, $\beta$=150, γ=50)

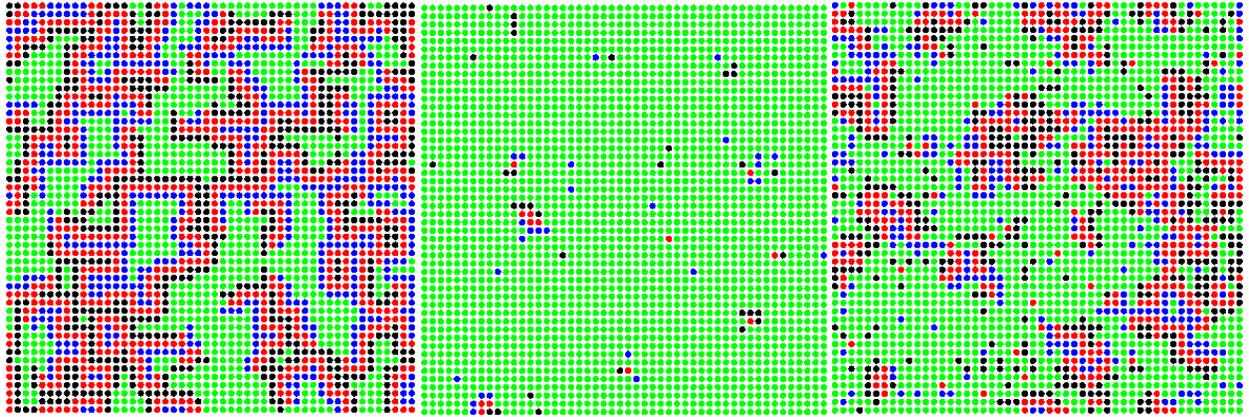

Figure 1 : Simulation results of changing participation rate and mutation rate
(Left : p = 1, μ = 0, Middle : p = 1, μ = 0.01, Right : p = 0.1, μ = 0.01)

We define coexistence value φ to measure how equally four types of agents are spread throughout the society, which is equal to $\sqrt[4]{n_c \times n_d \times n_l \times n_p}$. By using this value, we can easily conclude that condition, p = 1, μ = 0, results in highest coexistence value.

Interesting fact is that, when mutation rate is small cooperator is dominant when participation rate is the smallest. For example, μ = 0.0001, 0.001 number of cooperators is the largest when p = 0.001.
But μ becomes 0.01, cooperator is dominant when p = 0.01. When μ = 0.1, number of cooperator is the largest when p = 0.1. On the other hand, when p = 0.001, number of cooperator is the largest when μ = 0.001. As participation rate becomes higher, number of cooperator become larger when higher value of mutation rate is obtained.

Table 3 summarized our results by calculating average number of each agents and result of t-test. We can find that coexisting condition changes depending on participation rate and mutation rate. And Figure 2 shows that φ becomes larger when more agents mutate with small participation rate. To compare with other networks, number of agent is set as 1,000. Null hypothesis is p = 1, μ = 0 for the case of μ = 0. And for each participation rate, μ = 0 is set as null hypothesis. (+: P < 0.05, *: P < 0.01).

| Participation rate | Mutation rate | Cooperator Mean | t-value | Defector Mean | t-value | Loner Mean | t-value | Punisher Mean | t-value | Coexistence Value φ |
|---|---|---|---|---|---|---|---|---|---|---|
| 0.001 | 0.0001 | 277.0 | -6.6* | 211.1 | -4.2* | 310.5 | 8.4* | 201.5 | -4.6* | 245.9 |
| 0.001 | 0.001 | **377.9** | 18.4* | 217.7 | -2.9* | 195.1 | -9.5* | 209.3 | -4.5* | 240.7 |
| 0.001 | 0.01 | 268.4 | -16.7* | 234.1 | 1.6 | 258.7 | 13.4* | 238.7 | 4.7* | 249.6 |
| 0.001 | 0.1 | 251.4 | -24.3* | 247.0 | 7.6* | 248.9 | 8.3* | 252.3 | 10.4* | **249.9** |
| 0.01 | 0.0001 | 162.4 | -15.1* | 134.8 | -11.0* | 567.2 | 13.9* | 135.7 | -12.3* | 202.6 |

| 0.01 | 0.001 | 171.6 | -16.6* | 145.9 | -12.1* | 529.6 | 16.1* | 153.0 | -9.3* | 212.2 |
| 0.01 | 0.01 | **380.3** | 32.6* | 214.6 | 5.4* | 196.5 | -45.8* | 208.6 | 3.6* | 240.5 |
| 0.01 | 0.1 | 271.1 | -0.1 | 236.7 | 16.5* | 252.7 | -27.4* | 238.9 | 16.5* | **249.5** |
| 0.1 | 0.0001 | 142.0 | -23.9* | 118.0 | -21.7* | 622.0 | 21.8* | 118.0 | -15.3* | 187.3 |
| 0.1 | 0.001 | 3.7 | -377.8* | 10.0 | -221.7* | 977.7 | 302.6* | 8.7 | -202.8* | 23.6 |
| 0.1 | 0.01 | 140.0 | -19.2* | 132.9 | -15.2* | 590.6 | 18.6* | 136.5 | -12.7* | 196.8 |
| 0.1 | 0.1 | **375.6** | 28.3* | 205.1 | 0.4 | 202.6 | -41.4* | 216.3 | 9.7* | **241.1** |
| 1 | 0.0001 | 100.0 | -30.3* | 84.6 | -23.8* | 731.0 | 28.4* | 84.2 | -25.3* | 151.1 |
| 1 | 0.001 | 1.4 | -408.7* | 2.3 | -287.5* | 994.7 | 383.1* | 1.6 | -405.0* | 8.5 |
| 1 | 0.01 | 4.7 | -350.0* | 12.0 | -163.1* | 973.4 | 285.9* | 10.2 | -197.0* | 27.3 |
| 1 | 0.1 | **212.1** | -7.7* | 174.3 | -5.4* | 445.6 | 7.5* | 168.3 | -5.6* | **229.5** |
| 0.001 | 0 | 315.5 | 7.8* | 230.0 | 3.9* | 226.1 | -12.0* | 228.5 | 6.0* | 247.4 |
| 0.01 | 0 | 271.5 | 0.4 | 200.1 | -0.2 | 330.7 | -0.3 | 197.8 | 0.6 | 244.1 |
| 0.1 | 0 | 268.2 | -0.2 | 204.1 | 0.6 | 336.0 | 0.1 | 191.9 | -0.6 | 243.7 |
| 1 | 0 | 269.1 | 0.0 | 201.2 | 0.0 | 335.2 | 0.0 | 194.4 | 0.0 | 243.7 |

Table 3 : Different coexisting condition by changing participation and mutation rate (Moore, 1000)

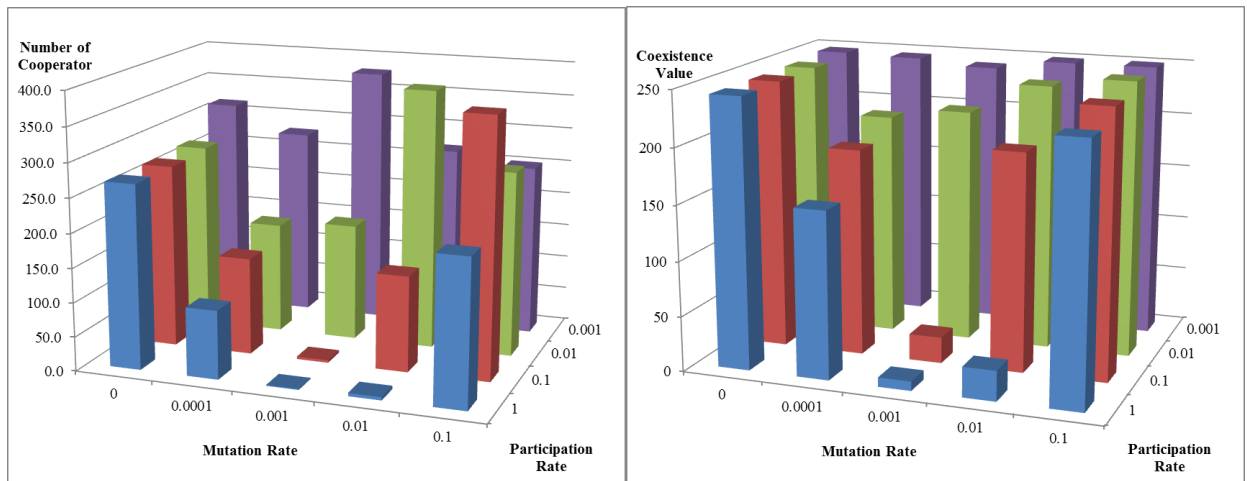

Figure 2 : Coexistence value (left) and Number of cooperator (right) of each cases

### 3.3   Network degree

We conducted same simulation by changing network into von-Neumann neighborhood, which has network degree of 4. Interestingly, we found that number of loner increase than the previous simulation with Moore neighborhood. The main reason loner becomes dominant is because degree of neighborhood is decreased to 4, which means that chance of having cooperator neighborhood is decreased. As agent doesn't have any cooperator neighborhood, they can't choose to be cooperator, defector, and loner because their payoff function becomes negative or 0. But in our payoff function, loner get non-zero payoff for any circumstances so agent with no cooperator neighbor will become loner eventually. On the other hand, by simulating network with degree = 16, larger number of cooperator and coexistence value is observed. Null hypothesis here is the same case of Moore neighborhood. (+: $P < 0.05$, *: $P < 0.01$).

| Participation rate | Mutation rate | Cooperator | | Defector | | Loner | | Punisher | | Coexistence Value φ |
|---|---|---|---|---|---|---|---|---|---|---|
| | | Mean | t-value | Mean | t-value | Mean | t-value | Mean | t-value | |
| 0.001 | 0.0001 | 51.1 | -116.7* | 94.8 | -47.0* | **756.7** | 72.4* | 97.5 | -39.5* | 137.5 |
| 0.001 | 0.001 | 206.2 | -66.8* | 234.7 | 5.2* | 327.9 | 42.3* | 231.3 | 6.6* | 246.1 |
| 0.001 | 0.01 | 243.4 | -9.0* | 245.0 | 4.0* | 261.3 | 1.0 | 250.1 | 4.9* | 249.8 |
| 0.001 | 0.1 | **248.6** | -1.1 | 249.9 | 1.3 | 248.6 | -0.1 | 252.6 | 0.1 | **249.9** |
| 0.01 | 0.0001 | 7.5 | -211.7* | 13.5 | -125.8* | **965.4** | 153.3* | 13.6 | -113.7* | 34.0 |

| | | | | | | | | | |
|---|---|---|---|---|---|---|---|---|---|
| 0.01 | 0.001 | 34.9 | -111.0* | 73.8 | -33.6* | 816.5 | 65.0* | 74.9 | -39.4* | 112.0 |
| 0.01 | 0.01 | 205.0 | -59.6* | 233.5 | 6.3* | 328.9 | 36.4* | 231.9 | 8.5* | 245.8 |
| 0.01 | 0.1 | **249.5** | -9.2* | 245.8 | 4.0* | 255.5 | 1.0 | 248.9 | 3.8* | **249.9** |
| 0.1 | 0.0001 | 2.8 | -244.4* | 4.4 | -143.2* | **988.3** | 170.8* | 4.4 | -139.3* | 15.3 |
| 0.1 | 0.001 | 2.9 | -2.1+ | 8.9 | -1.6 | 979.7 | 1.5 | 8.7 | 0.0 | 21.7 |
| 0.1 | 0.01 | 34.3 | -84.7* | 75.9 | -29.4* | 812.8 | 52.3* | 77.9 | -26.9* | 113.3 |
| 0.1 | 0.1 | **215.5** | -52.3* | 230.3 | 8.1* | 317.0 | 30.6* | 237.1 | 9.5* | **247.2** |
| 1 | 0.0001 | 1.6 | -257.6* | 2.1 | -172.2* | 993.6 | 182.1* | 2.6 | -139.1* | 9.7 |
| 1 | 0.001 | 0.3 | -10.6* | 1.0 | -6.0* | **997.6** | 5.3* | 1.0 | -2.0+ | 4.0 |
| 1 | 0.01 | 3.6 | -2.1+ | 8.8 | -3.3* | 978.4 | 2.5* | 9.2 | -1.2 | 23.1 |
| 1 | 0.1 | **41.7** | -112.1* | 89.0 | -42.0* | 785.3 | 75.0* | 85.2 | -36.7* | **125.6** |
| 0.001 | 0 | 106.8 | -69.5* | 169.0 | -13.0* | 558.4 | 37.8* | 165.8 | -17.2* | 202.2 |
| 0.01 | 0 | 67.6 | -109.7* | 90.3 | -48.9* | 753.4 | 69.4* | 88.8 | -49.3* | 142.1 |
| 0.1 | 0 | 63.0 | -92.6* | 84.2 | -45.4* | 768.8 | 57.7* | 84.1 | -37.8* | 136.1 |
| 1 | 0 | 72.1 | -92.2* | 94.3 | -39.2* | 737.3 | 54.9* | 95.5 | -35.9* | 147.9 |

Table 4 : Simulation result with von-Neumann neighborhood lattice

| Participation rate | Mutation rate | Cooperator | | Defector | | Loner | | Punisher | | Coexistence Value |
|---|---|---|---|---|---|---|---|---|---|---|
| | | Mean | t-value | Mean | t-value | Mean | t-value | Mean | t-value | |
| 0.001 | 0.0001 | 263.6 | -3.4* | 225.3 | 3.3* | 273.6 | -7.2* | 237.4 | 8.5* | 249.2 |
| 0.001 | 0.001 | **422.4** | 14.5* | 219.1 | 0.4 | 131.7 | -25.5* | 226.9 | 5.3* | 229.3 |
| 0.001 | 0.01 | 292.5 | 9.6* | 238.7 | 2.1+ | 234.4 | -9.6* | 234.9 | -1.5 | 249.0 |
| 0.001 | 0.1 | 256.4 | 2.4* | 245.2 | -0.7 | 247.5 | -0.6* | 251.3 | -0.5 | **250.1** |
| 0.01 | 0.0001 | 172.4 | 2.0 | 145.0 | 2.2+ | 538.6 | -2.3+ | 144.0 | 2.2+ | 209.9 |
| 0.01 | 0.001 | 262.2 | 21.3* | 223.8 | 14.4* | 276.9 | -30.1* | 237.0 | 15.6* | **249.1** |
| 0.01 | 0.01 | **419.1** | 16.9* | 226.7 | 3.7* | 131.7 | -33.4* | 221.8 | 4.0* | 229.5 |
| 0.01 | 0.1 | 290.6 | 8.2* | 241.3 | 1.7+ | 227.7 | -13.0* | 240.6 | 0.9 | 249.0 |
| 0.1 | 0.0001 | 153.5 | 2.0 | 126.1 | 1.8+ | 596.8 | -1.7+ | 123.6 | 1.2 | 194.4 |
| 0.1 | 0.001 | 155.6 | 34.7* | 131.0 | 34.1* | 578.8 | -36.3* | 134.7 | 33.1* | 199.6 |
| 0.1 | 0.01 | 264.1 | 25.6* | 227.6 | 20.1* | 277.6 | -40.1* | 229.8 | 18.2* | **248.8** |
| 0.1 | 0.1 | **414.6** | 14.3* | 218.9 | 4.5* | 140.6 | -27.5* | 227.1 | 2.9* | 232.0 |
| 1 | 0.0001 | 138.7 | 10.6* | 116.3 | 9.7* | 630.2 | -10.7* | 114.8 | 10.0* | 184.8 |
| 1 | 0.001 | 127.1 | 28.0* | 106.7 | 31.4* | 660.6 | -31.4* | 105.6 | 33.5* | 175.4 |
| 1 | 0.01 | 142.7 | 25.1* | 120.9 | 31.3* | 611.4 | -29.8* | 124.6 | 29.2* | 190.4 |
| 1 | 0.1 | **267.3** | 13.8* | 218.4 | 9.3* | 287.2 | -20.1* | 226.5 | 12.1* | **248.2** |
| 0.001 | 0 | 283.3 | -8.4* | 270.3 | 5.5* | 166.7 | -12.8* | 279.7 | 7.9* | 244.5 |
| 0.01 | 0 | 197.2 | -18.7* | 161.5 | -12.0* | 476.0 | 15.0* | 165.2 | -10.2* | 223.7 |
| 0.1 | 0 | 198.5 | -14.5* | 164.7 | -11.0* | 472.5 | 12.3* | 164.3 | -9.0* | 224.5 |
| 1 | 0 | 192.1 | -24.9* | 159.8 | -16.5* | 486.5 | 22.0* | 161.6 | -14.1* | 221.6 |

Table 5 : Simulation result with lattice, degree=16

### 3.4 Network Structure

We conducted same simulation by applying complex network structures. Cellular, Core-periphery, Erdős-Rényi, Small-world, Scale-free network with same number of node and edge are used. Since network degree of each node diverse in complex network, random seed which decides agent's initial distribution over the network becomes critical here. For example, if node with degree 100 is chosen as a cooperator, its 100 neighbors will get benefit by it, so overall agent dynamics will reflect that positive effect. In that case, much more cooperators will be survived to the end. Figure 3 shows how different the stable states of agent will change, depending on the initial condition by changing random seed (Network information: Erdős-Rényi random network with p = 0.01, μ = 0).

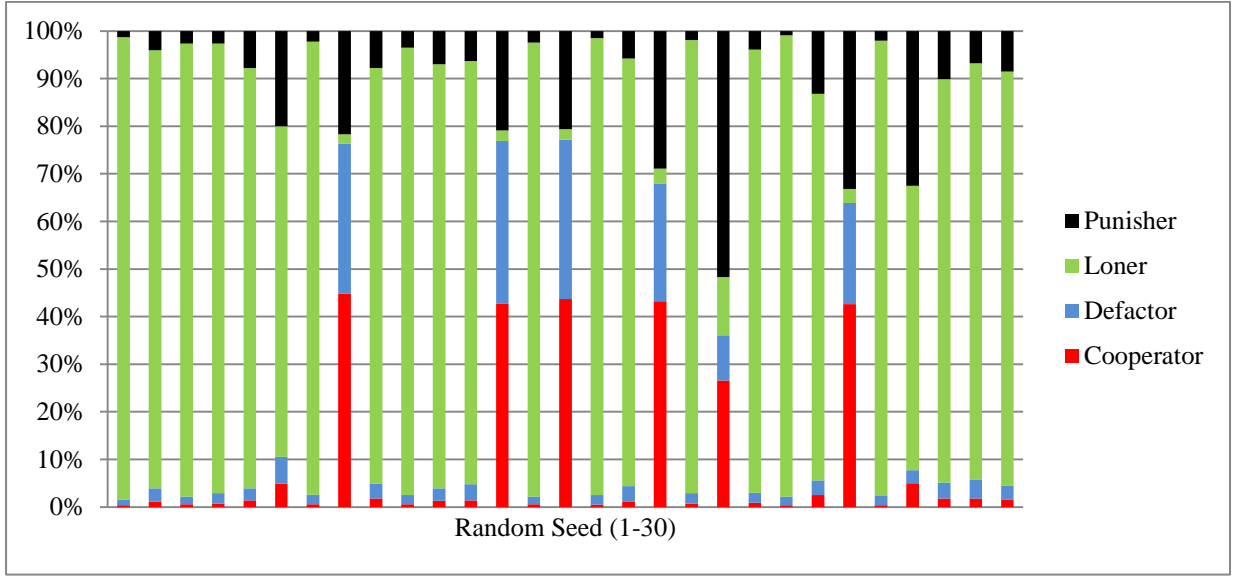

Figure 3 : Different agent behavior by changing random seed

We used meta-network assessment and analysis tool ORA(Carley, Kathleen M et al., 2012) to generate six different stylized networks. To restrict our focus into network structure, we fixed the number of nodes to 1,000 and made number of edge as close as 8,000. We didn't remove any surplus edges not to destruct network characteristic. Table 6 shows detail explanation of each network.

| Network | Number of Node | Number of Edge | Other information |
|---|---|---|---|
| Grid(Moore) | 1,000, 2,500 | | Used Grid space provided by Repast |
| Grid(von-Neumann) | 1,000, 2,500 | | Used Grid space provided by Repast |
| Lattice | 1,000 | 8,000 | Directed Degree of 8 - Number of neighbor : 16 |
| Cellular | 1,000 | 7,995 | Number of Cells : 52, Inner Density : 0.40, Outer Density : 0.10 |
| Core-Periphery | 1,000 | 8,003 | Proportion of core nodes : 0.13, Density of core nodes : 0.50 |
| Erdős-Rényi, | 1,000 | 8,000 | Total density : 0.008 (symmetric) |
| Scale-Free | 1,000 | 8,004 | Probability of node connecting to core : 0.008<br>Initial node count(core) : 40, Initial density(core) : 0.01<br>Number of node connected by edge : 891 (109 nodes has degree 0) |
| Small-World | 1,000 | 8,005 | Number of neighbors : 8, Probability of removing neighbor : 0.05<br>Probability of adding for neighbor : 0.055, Power law exponent : 0.055 |
| Network made by ORA, which is directed graph, but we used it as un-directed. | | | |

Table 6 : Specific Network Information

We found the result of networks having heterogeneous degree is more fluctuated than homogeneous network by observing standard deviation. Figure 4, 5 shows the mean and standard deviation of each agent type. In the condition of p=0.1, μ=0.001, Erdős-Rényi and Scale-free network has the biggest standard deviation because of their diversity of node degree. Cooperator is not appeared in lattice network since stable structure is broken because of mutation rate (Section 3.2). However by changing graph structure, we can find coexisting behavior in three types of networks – Erdős-Rényi, Scale-free, Small-world. Simulation results by varying network structures are in APPENDIX.

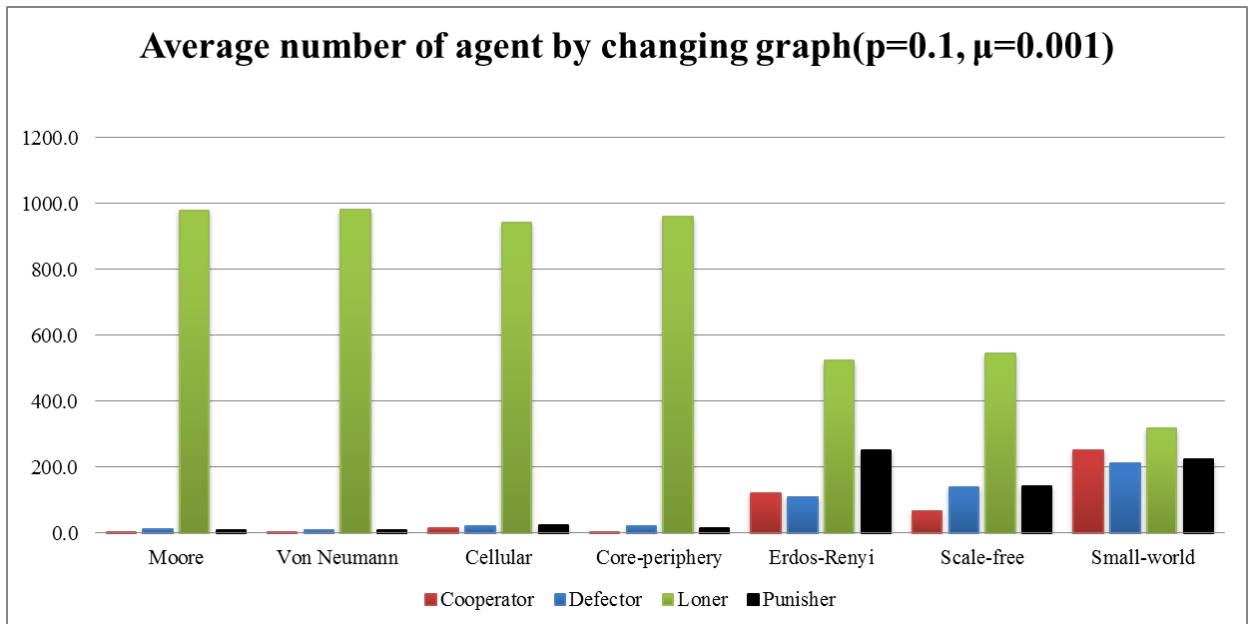

Figure 4 : Average number of agent by changing network structure (p=0.1, μ=0.001)

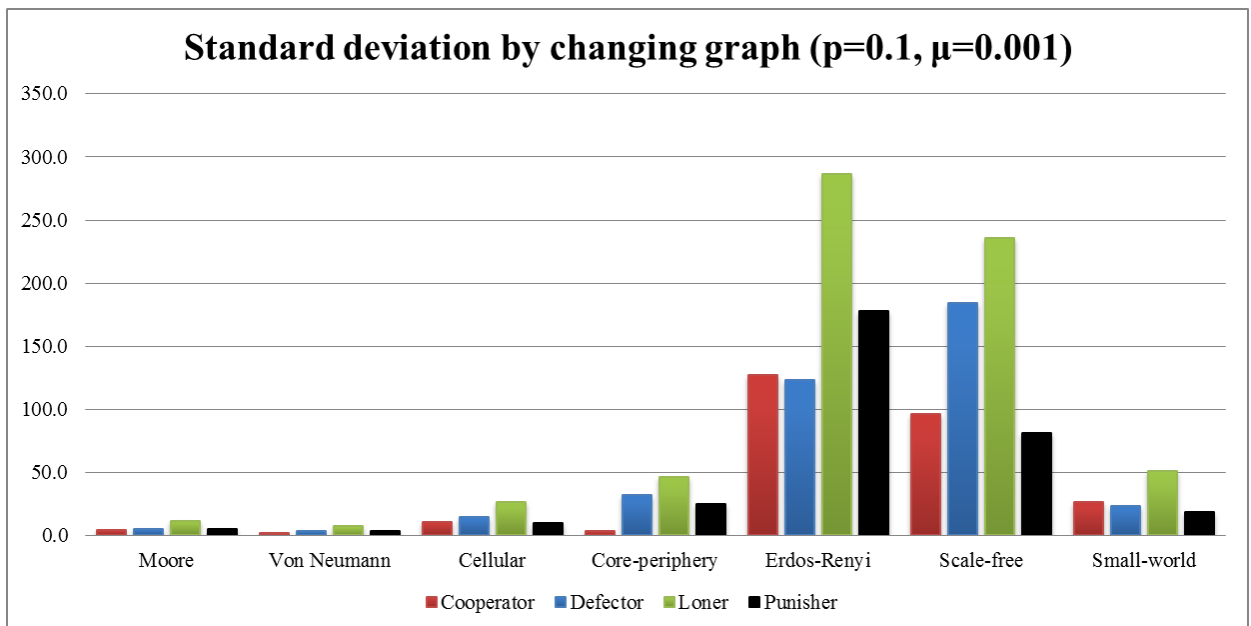

Figure 5 : Standard deviation of agent by changing network structure(p=0.1, μ=0.001)

### 3.5 Limitation of our simulation model

In our model, we used pre-defined participation rate and mutation rate value, to verify that those parameters affect our model stochastically. But in the real world agent can decide themselves whether joining for certain game on each time, it is not decided by force. And agents can choose different strategy by their own utility function, not just combining the payoff depending on their neighborhood. Making agents payoff heterogeneous depends on each agent's background will be needed to show organizational behavior better. In order to use self-motivated agent model in the real world, calibration depending on the situation is needed. Diverse type of real world dataset will be needed to explain the society.

Second, we assume that each agent knows current state of its neighbors. But in real life, information imbalance makes not everyone can decide their decision in optimal way. Also more accurate initial condition will be considered.

Last, stochastic difference can be occurred, if sequence of applying participation rate and mutation rate is changed. Although we reflect participation and mutation is independent, our result shows that there is a correlation between two variables.

## 4    CONCLUSIONS

In this paper, we studied the behavior of 4 different types of self-motivated agents playing N-person prisoner's dilemma game. By using agent based modeling, we can reflect the agent's mutation and participation in stochastic way. And we could observe different behavior by changing network structure.

We found the impact of participation rate and mutation rate on agent society. Degree of network is also very critical factor changing population of each agent type. And we found that structure of network is important in that each agent affect society different according to heterogeneity. Besides some cases especially in Erdős-Rényi, network, we found the phenomenon that initial location of agent affect their coexisting condition critically.

For further work, we will compare how self-motivated agent behaves different with agents in traditional evolutionary games. By comparing different coexisting condition, we can analyze which model explains our human nature and society better. Also, we will put different utility function while calculating payoff for emphasizing heterogeneity of agent. Last, we will prove how participation rate affects with mutation rate in analytical way.

## A    APPENDIX

Table 5 to 8 is the simulation result of agents' behavior in different complex networks. We can find that participation – mutation relation is also followed in complex networks. Null hypothesis is $p = 1$, $\mu = 0$ for the case of $\mu = 0$. And for each participation rate, $\mu = 0$ is set as null hypothesis. (+: $P < 0.05$, *: $P < 0.01$).

| Participation rate | Mutation rate | Cooperator Mean | t-value | Defector Mean | t-value | Loner Mean | t-value | Punisher Mean | t-value | Coexistence Value φ |
|---|---|---|---|---|---|---|---|---|---|---|
| 0.001 | 0.0001 | 115.1 | -24.5* | 101.8 | -7.7* | 527.2 | 38.3* | 255.9 | -39.2* | 199.4 |
| 0.001 | 0.001 | **417.2** | 72.4* | 153.8 | 8.6* | 140.2 | -14.0* | 288.6 | -63.4* | 225.8 |
| 0.001 | 0.01 | 299.1 | 34.2* | 230.4 | 45.4* | 237.3 | 27.8* | 233.3 | -92.6* | 248.5 |
| 0.001 | 0.1 | 254.2 | 15.2* | 249.2 | 47.4* | 246.7 | 41.2* | 249.7 | -95.4* | **250.0** |
| 0.01 | 0.0001 | 9.8 | 3.0* | 20.4 | 8.3* | 949.9 | -9.2* | 19.8 | 7.1* | 44.1 |
| 0.01 | 0.001 | 145.4 | 23.1* | 139.7 | 21.6* | 469.9 | -43.5* | 245.1 | 34.1* | 219.9 |
| 0.01 | 0.01 | **413.0** | 119.7* | 156.2 | 55.8* | 140.3 | -424.3* | 291.1 | 61.1* | 226.6 |
| 0.01 | 0.1 | 293.5 | 110.2* | 237.2 | 84.6* | 229.2 | -328.1* | 240.4 | 110.9* | **248.9** |
| 0.1 | 0.0001 | 2.8 | -6.9* | 3.1 | -2.5* | 990.6 | 3.7* | 3.5 | 0.0 | 13.2 |
| 0.1 | 0.001 | 13.1 | 2.3* | 21.2 | 6.2* | 942.0 | -8.6* | 23.4 | 10.9* | 49.7 |
| 0.1 | 0.01 | 135.9 | 28.5* | 123.3 | 27.5* | 495.0 | -46.4* | 245.2 | 33.4* | 212.3 |
| 0.1 | 0.1 | **416.5** | 133.6* | 163.8 | 60.0* | 144.9 | -336.0* | 276.4 | 79.6* | **228.7** |
| 1 | 0.0001 | 1.5 | -4.9* | 1.1 | -3.4* | 996.9 | 4.8* | 0.5 | -7.1* | 5.3 |
| 1 | 0.001 | 7.4 | 1.5 | 5.8 | 2.2+ | 982.7 | -2.2+ | 4.1 | 3.0* | 20.3 |
| 1 | 0.01 | 14.1 | 4.7* | 21.6 | 7.5* | 941.3 | -10.0* | 23.3 | 9.7* | 50.8 |
| 1 | 0.1 | **134.0** | 26.1* | 131.4 | 20.8* | 513.3 | -44.2* | 219.2 | 36.5* | **211.0** |
| 0.001 | 0 | 216.9 | 33.3* | 138.3 | 20.7* | 165.9 | -119.7* | 478.9 | 58.4* | 221.0 |
| 0.01 | 0 | 5.8 | 0.6 | 3.7 | 1.0 | 988.1 | -0.8 | 2.5 | 0.9 | 15.1 |
| 0.1 | 0 | 8.5 | 1.7+ | 4.8 | 1.8+ | 983.1 | -1.8+ | 3.5 | 1.9+ | 19.4 |
| 1 | 0 | 4.9 | 0.0 | 2.8 | 0.0 | 990.3 | 0.0 | 2.0 | 0.0 | 12.7 |

Table 7 : Simulation result with Cellular Network

| Participation rate | Mutation rate | Cooperator | | Defector | | Loner | | Punisher | | Coexistence Value φ |
|---|---|---|---|---|---|---|---|---|---|---|
| | | Mean | t-value | Mean | t-value | Mean | t-value | Mean | t-value | |
| 0.001 | 0.0001 | 26.4 | -0.1 | 38.9 | -10.4* | 834.3 | 14.2* | 100.3 | -20.1* | 96.3 |
| 0.001 | 0.001 | 180.3 | 64.3* | 206.1 | 39.4* | 399.2 | -60.6* | 214.6 | 5.4* | 237.5 |
| 0.001 | 0.01 | 240.9 | 101.4* | 236.4 | 59.7* | 288.8 | -163.0* | 234.7 | 12.1* | 249.3 |
| 0.001 | 0.1 | **251.2** | 104.6* | 246.7 | 64.2* | 249.7 | -189.3* | 252.7 | 24.5* | **250.1** |
| 0.01 | 0.0001 | 2.9 | 8.2* | 26.8 | 3.7* | 955.2 | -5.5* | 15.1 | 5.3* | 32.5 |
| 0.01 | 0.001 | 26.3 | 17.3* | 38.6 | 6.6* | 833.8 | -17.0* | 101.3 | 18.7* | 96.2 |
| 0.01 | 0.01 | 178.4 | 79.2* | 211.3 | 85.4* | 394.4 | -137.1* | 216.5 | 79.1* | 238.2 |
| 0.01 | 0.1 | **238.8** | 105.5* | 240.0 | 93.1* | 280.6 | -270.0* | 240.8 | 94.4* | **249.4** |
| 0.1 | 0.0001 | 0.2 | 2.5* | 2.3 | 1.2 | 996.9 | -1.5 | 0.6 | 2.1+ | 4.2 |
| 0.1 | 0.001 | 3.5 | 5.2* | 20.7 | 3.5* | 960.6 | -4.7* | 14.8 | 3.3* | 32.0 |
| 0.1 | 0.01 | 26.0 | 21.5* | 33.9 | 18.8* | 835.8 | -24.2* | 104.7 | 18.8* | 93.7 |
| 0.1 | 0.1 | **186.3** | 97.3* | 211.9 | 86.8* | 387.9 | -206.2* | 214.9 | 94.3* | **239.5** |
| 1 | 0.0001 | 0.0 | 0.0 | 0.0 | 0.0 | 1000.0 | -1.0 | 0.0 | 1.0 | 0.0 |
| 1 | 0.001 | 0.4 | 3.3* | 0.4 | 3.6* | 999.1 | -6.1* | 0.3 | 3.2* | 2.4 |
| 1 | 0.01 | 3.3 | 7.8* | 20.5 | 2.9* | 958.0 | -4.6* | 19.1 | 4.0* | 33.4 |
| 1 | 0.1 | **28.3** | 18.2* | 43.0 | 12.3* | 809.4 | -25.5* | 120.1 | 27.6* | **104.3** |
| 0.001 | 0 | 26.6 | 18.2* | 104.5 | 22.9* | 668.4 | -37.7* | 200.5 | 55.0* | 138.9 |
| 0.01 | 0 | 0.0 | 0.0 | 0.0 | 0.0 | 1000.0 | 0.0 | 0.0 | 0.0 | 0.0 |
| 0.1 | 0 | 0.0 | 0.0 | 0.0 | 0.0 | 1000.0 | 0.0 | 0.0 | 0.0 | 0.0 |
| 1 | 0 | 0.0 | 0.0 | 0.0 | 0.0 | 1000.0 | 0.0 | 0.0 | 0.0 | 0.0 |

Table 8 : Simulation result with Core-Periphery Network

| Participation rate | Mutation rate | Cooperator | | Defector | | Loner | | Punisher | | Coexistence Value φ |
|---|---|---|---|---|---|---|---|---|---|---|
| | | Mean | t-value | Mean | t-value | Mean | t-value | Mean | t-value | |
| 0.001 | 0.0001 | 222.0 | -14.6* | 55.2 | -27.7* | 181.7 | 8.7* | 541.1 | 24.9* | 186.3 |
| 0.001 | 0.001 | **422.2** | -11.9* | 193.0 | -0.7 | 136.1 | 46.6* | 248.8 | -20.5* | 229.2 |
| 0.001 | 0.01 | 286.9 | -76.2* | 239.1 | 20.2* | 239.2 | 91.3* | 235.5 | -36.2* | 249.3 |
| 0.001 | 0.1 | 252.7 | -86.2* | 247.1 | 20.4* | 247.2 | 94.8* | 252.6 | -33.7* | **249.9** |
| 0.01 | 0.0001 | 127.5 | 1.3 | 80.8 | 0.4 | 530.8 | -3.2* | 261.2 | 4.8* | 194.4 |
| 0.01 | 0.001 | 131.6 | 8.6* | 33.7 | -35.9* | 332.9 | -44.9* | 501.6 | 69.1* | 164.9 |
| 0.01 | 0.01 | **416.6** | 136.2* | 203.5 | 41.4* | 138.7 | -320.3* | 241.8 | 30.0* | 230.9 |
| 0.01 | 0.1 | 280.5 | 75.9* | 241.9 | 67.8* | 233.4 | -215.2* | 244.1 | 48.2* | **249.3** |
| 0.1 | 0.0001 | 0.3 | -572.7* | 1.3 | -158.5* | 996.7 | 233.0* | 1.5 | -128.2* | 4.9 |
| 0.1 | 0.001 | 120.6 | 2.8* | 107.6 | 3.1* | 522.1 | -6.6* | 249.3 | 6.4* | 202.8 |
| 0.1 | 0.01 | 135.0 | 21.7* | 36.7 | -0.4 | 325.1 | -66.3* | 504.1 | 90.7* | 168.8 |
| 0.1 | 0.1 | **412.3** | 146.8* | 204.0 | 51.7* | 145.3 | -354.8* | 240.7 | 54.5* | **232.9** |
| 1 | 0.0001 | 0.0 | -899.0* | 0.1 | -253.4* | 999.7 | 373.2* | 0.2 | -184.6* | 0.9 |
| 1 | 0.001 | 0.2 | -379.0* | 0.8 | -145.5* | 997.8 | 189.9* | 1.1 | -88.3* | 3.7 |
| 1 | 0.01 | 132.1 | 7.7* | 450.2 | 34.6* | 372.1 | -23.9* | 45.9 | 4.6* | **178.5** |
| 1 | 0.1 | **154.6** | 25.1* | 41.2 | 13.7* | 288.9 | -61.5* | 512.6 | 67.8* | 175.2 |
| 0.001 | 0 | 454.6 | 117.4* | 194.9 | 18.3* | 32.2 | -817.2* | 318.4 | 26.0* | 173.6 |
| 0.01 | 0 | 93.8 | 2.2+ | 73.4 | 2.5* | 716.3 | -3.2* | 116.5 | 4.2* | 154.8 |
| 0.1 | 0 | 56.2 | 1.0 | 37.1 | 0.6 | 864.0 | -0.9 | 42.6 | 1.1 | 93.6 |
| 1 | 0 | 30.0 | 0.0 | 26.6 | 0.0 | 921.4 | 0.0 | 22.0 | 0.0 | 63.4 |

Table 9 : Simulation result with Erdős-Rényi, Network

| Participation rate | Mutation rate | Cooperator | | Defector | | Loner | | Punisher | | Coexistence Value φ |
|---|---|---|---|---|---|---|---|---|---|---|
| | | Mean | t-value | Mean | t-value | Mean | t-value | Mean | t-value | |
| 0.001 | 0.0001 | 93.0 | -63.5* | 31.6 | -100.1* | 341.3 | 28.2* | 425.0 | 9.0* | 143.7 |
| 0.001 | 0.001 | **338.5** | 2.8* | 176.3 | 6.4* | 136.1 | 43.5* | 240.0 | -29.8* | 210.1 |
| 0.001 | 0.01 | 252.5 | -24.7* | 213.3 | 22.4* | 213.0 | 63.2* | 212.3 | -66.5* | 222.1 |
| 0.001 | 0.1 | 222.3 | -43.0* | 217.4 | 27.2* | 225.0 | 64.1* | 224.6 | -51.7* | **222.3** |
| 0.01 | 0.0001 | 100.2 | 3.8* | 96.0 | 3.1* | 476.3 | -6.7* | 218.4 | 7.8* | 177.9 |
| 0.01 | 0.001 | 84.6 | 26.7* | 28.6 | 24.5* | 383.2 | -52.7* | 394.7 | 55.4* | 138.3 |
| 0.01 | 0.01 | **339.2** | 142.8* | 176.5 | 48.7* | 139.3 | -452.3* | 236.2 | 44.2* | 210.7 |
| 0.01 | 0.1 | 248.8 | 92.6* | 211.4 | 75.4* | 215.9 | -214.7* | 213.6 | 84.6* | **221.9** |
| 0.1 | 0.0001 | 50.8 | 2.6* | 100.5 | 3.0* | 713.1 | -3.0* | 26.4 | 2.5* | 99.0 |

| | | | | | | | | | |
|---|---|---|---|---|---|---|---|---|---|
| 0.1 | 0.001 | 65.8 | 3.7* | 140.2 | 4.2* | 543.4 | -8.1* | 141.7 | 9.6* | 163.2 |
| 0.1 | 0.01 | 84.7 | 23.1* | 29.3 | 26.8* | 381.8 | -61.0* | 394.9 | 73.4* | 139.1 |
| 0.1 | 0.1 | **337.1** | 121.8* | 184.9 | 52.8* | 138.9 | -435.5* | 228.4 | 48.9* | **210.9** |
| 1 | 0.0001 | 82.9 | 3.4* | 136.4 | 3.7* | 631.9 | -3.7* | 39.8 | 2.9* | 129.9 |
| 1 | 0.001 | **113.5** | 4.6* | 226.9 | 5.6* | 504.8 | -5.5* | 45.7 | 4.2* | 156.1 |
| 1 | 0.01 | 107.9 | 6.0* | 247.0 | 6.2* | 380.9 | -11.0* | 152.6 | 5.1* | **198.4** |
| 1 | 0.1 | 94.9 | 24.8* | 32.2 | 22.0* | 355.0 | -53.1* | 410.4 | 57.8* | 145.2 |
| 0.001 | 0 | 331.3 | 46.7* | 159.2 | 19.3* | 44.1 | -354.4* | 356.3 | 27.5* | 169.7 |
| 0.01 | 0 | 17.8 | 1.4 | 9.6 | 1.4 | 830.5 | -1.6 | 33.2 | 1.6 | 46.6 |
| 0.1 | 0 | 0.0 | 0.0 | 0.0 | 0.0 | 891.0 | 0.0 | 0.0 | 0.0 | 0.0 |
| 1 | 0 | 0.0 | 0.0 | 0.0 | 0.0 | 891.0 | 0.0 | 0.0 | 0.0 | 0.0 |

Table 10 : Simulation result with Scale-Free Network

| Participation rate | Mutation rate | Cooperator | | Defector | | Loner | | Punisher | | Coexistence Value φ |
|---|---|---|---|---|---|---|---|---|---|---|
| | | Mean | t-value | Mean | t-value | Mean | t-value | Mean | t-value | |
| 0.001 | 0.0001 | 286.1 | -9.3* | 169.3 | -11.8* | 217.6 | 16.6* | 327.0 | -4.9* | 242.3 |
| 0.001 | 0.001 | **416.7** | 25.3* | 189.0 | -12.7* | 136.1 | 17.0* | 258.2 | -24.8* | 229.4 |
| 0.001 | 0.01 | 291.6 | -12.7* | 238.1 | 7.0* | 234.6 | 55.0* | 235.9 | -42.0* | 249.0 |
| 0.001 | 0.1 | 254.5 | -31.9* | 249.2 | 12.5* | 244.3 | 81.8* | 250.6 | -41.0* | **249.6** |
| 0.01 | 0.0001 | 248.4 | 3.9* | 212.9 | 1.1 | 324.0 | -3.5* | 214.5 | 4.1* | 246.2 |
| 0.01 | 0.001 | 275.5 | 9.5* | 155.4 | -12.6* | 240.1 | -16.1* | 328.7 | 27.9* | 241.1 |
| 0.01 | 0.01 | **422.9** | 60.3* | 195.8 | -3.8* | 134.2 | -112.7* | 247.6 | 11.9* | 229.0 |
| 0.01 | 0.1 | 284.1 | 27.3* | 241.7 | 13.4* | 229.8 | -63.4* | 245.0 | 20.0* | **249.3** |
| 0.1 | 0.0001 | 242.7 | 3.8* | 212.3 | 2.0+ | 338.8 | -3.3* | 205.9 | 2.6* | 244.8 |
| 0.1 | 0.001 | 249.2 | 5.8* | 210.2 | 1.9+ | 316.8 | -6.7* | 223.9 | 7.8* | **246.9** |
| 0.1 | 0.01 | 280.4 | 14.7* | 168.3 | -5.9* | 232.5 | -19.6* | 319.2 | 24.1* | 243.3 |
| 0.1 | 0.1 | **419.1** | 65.6* | 196.8 | -2.4* | 141.5 | -107.4* | 244.0 | 12.3* | 231.0 |
| 1 | 0.0001 | 233.1 | 2.2+ | 210.9 | 2.3+ | 359.3 | -1.8+ | 196.7 | 0.1 | 242.8 |
| 1 | 0.001 | 241.9 | 4.2* | 214.5 | 3.4* | 339.8 | -3.7* | 203.9 | 2.0+ | 244.9 |
| 1 | 0.01 | 248.9 | 4.7* | 216.8 | 3.4* | 314.8 | -5.3* | 219.5 | 4.8* | **247.1** |
| 1 | 0.1 | **296.9** | 15.0* | 180.0 | -3.8* | 221.1 | -22.1* | 301.4 | 19.3* | 244.3 |
| 0.001 | 0 | 327.1 | 27.4* | 222.5 | 3.0* | 103.8 | -67.7* | 346.6 | 19.5* | 226.2 |
| 0.01 | 0 | 223.7 | 0.8 | 207.9 | 1.3 | 370.1 | -1.1 | 198.3 | 0.6 | 241.7 |
| 0.1 | 0 | 221.3 | 0.0 | 202.3 | 0.0 | 378.4 | 0.0 | 197.9 | 0.0 | 240.6 |
| 1 | 0 | 219.5 | 0.0 | 200.1 | 0.0 | 384.0 | 0.0 | 196.3 | 0.0 | 239.9 |

Table 11 : Simulation result with Small-World Network

**AUTHOR BIOGRAPHIES**

**SUNDONG KIM** is a Master candidate at Department of Industrial and Systems Engineering, KAIST. His email address is -sdkim@kaist.ac.kr.

**JIN-JAE LEE** is a Ph.D. candidate at Department of Physics, KAIST. His email address is wwpjin@kaist.ac.kr